\documentclass[amsmath,aps,showpacs,preprint]{revtex4}
\usepackage{amsfonts}

\usepackage[dvipdfm]{graphicx}
\usepackage[dvipdfm]{rotating}
\usepackage{amsmath}
\usepackage{epsfig}
\usepackage{bm}

\begin{document}

\title{Diffusive and precessional spin dynamics in
a two-dimensional electron gas with disorder: a gauge theory view}
\author{I.V. Tokatly$^{1,2,4}$ and E. Ya. Sherman$^{3,4}$}
\affiliation{$^1$ European Theoretical Spectroscopy Facility (ETSF), Departamento de Fisica de
Materiales,  Universidad del Pa\'is Vasco UPV/EHU and Centro Mixto CSIC-UPV/EHU, San Sebastian, Spain \\
$^2$  Moscow Institute of Electronic Technology, Zelenograd, 124498 Russia \\
$^3$ Department of Physical Chemistry, Universidad del Pa\'is Vasco UPV/EHU,
48080 Bilbao, Spain\\
$^4$  IKERBASQUE Basque Foundation for Science, Alameda Urquijo
36-5, 48011, Bilbao, Bizkaia, Spain 
}

\begin{abstract}
We develop a gauge theory for diffusive and precessional spin dynamics in
two-dimensional electron gas with disorder. Our approach reveals a direct connections between the
absence of the equilibrium spin current and strong anisotropy in the spin
relaxation: both effects arise if the spin-orbit coupling is reduced to a
pure gauge $SU(2)$ field. In this case, by a gauge transformation in the form of a local $%
SU(2)$ rotation in the spin subspace the spin-orbit coupling can be removed. 
The resulting spin dynamics is exactly described in terms of two kinetic coefficients:
the spin diffusion and electron mobility. After the inverse
transformation, full diffusive and precessional spin density dynamics,
including the anisotropic spin relaxation, formation of stable spin
structures, and spin precession induced by a macroscopic current, is
restored. Explicit solutions of the spin evolution equations 
are found for the initially uniform spin 
density and for stable nonuniform structures. 
Our analysis demonstrates a universal relation between the spin
relaxation rate and spin diffusion coefficient.
\end{abstract}


\maketitle

\section{Introduction.}

Description of the spin dynamics of a 
two-dimensional electron gas is one of the most important
problems for fundamental and applied modern spintronics.\cite{Zutic04,Fabian07,Dyakonov08} Two mutually related problems in this field attract a great
deal of attention: spin current and spin relaxation. Spin currents describe
how the spin density pattern changes with time mainly due to the spin transfer
between different parts of the electron gas. Since the
spin dynamics of interest occurs usually in systems out of equilibrium,
spin relaxation becomes important and contributes strongly into the evolution
of spin density pattern.

The key for understanding these properties is the spin-orbit coupling,
making the orbital and spin degrees of freedom mutually dependent.
Spin-orbit coupling has many crucial influences
on the properties of the systems where it occurs: the typical 
examples are nuclei, elementary particles, atoms, and electrons in
solids. Spin-orbit coupling, in general, 
makes the spin a non-conserved quantity, thus leading to a spin
relaxation. It causes mutually dependent spin, charge, and mass flows in
solids and quantum liquids.\cite{Leurs08,Kleinert,Galitski} In addition, spin-orbit coupling leads to
a spin response to an external electric field, providing an ability of spin
manipulation by the electric field driving the dynamics in the orbital degrees
of freedom.\cite{Rashba03a}

The general techniques for calculation of spin relaxation and spin current
out of equilibrium are the classical or quantum Boltzmann-like equation for
the spin-density matrix \cite{Mishchenko04,Culcer07,Glazov07,Wu09,Bronold04} 
and nonequilibrium Green functions.\cite{Raimondi,Bleibaum}
In this approach, the description of the
electron dynamics takes into account possible relaxation processes due to
the presence of spin-orbit coupling, disorder, phonons, and
electron-electron collisions. The experimentally observable spin dynamics is
due to the spin-orbit (spin-momentum) coupling. At the equilibrium, 
the expectation value of the spin
current can be calculated directly. Surprisingly, such a direct calculation 
demonstrated that the spin current can
exist even in the equilibrium state of a two-dimensional electron gas with
spin-orbit coupling.\cite{Rashba03} This observation brought a puzzle for
the understanding of the basic phenomena in spin transport since the
equilibrium spin current is not related to any spatial
spin accumulation that can be seen experimentally.

On the other hand, the spin dynamics due to the spin-orbit coupling can be
understood in terms of a theory where the coupling is treated as a
non-Abelian gauge field, and the corresponding formalism can be applied \cite{Mineev92,Frolich93}. On
a single-electron scale, for example, for electrons in quantum dots, the
gauge transformation of the spin-orbit field was employed in Refs.\cite{Aleiner01,Levitov03}
and used for the analysis of experimental results on
spin manipulation by electric field in quantum dots in Ref.\cite{Nowack07}.
Other interesting theoretical examples of applications of the non-Abelian gauge
field approach for single-electron spin transport and electrons in quantum
dots, were found and studied (for example, Refs.\cite{Yang08}-\cite{Yang06}).

When applied to the two-dimensional electron gas, the approach based on a formal $SU(2)$ gauge
invariance of the spin-orbit Hamiltonian (i.~e. the symmetry with respect to local rotations in the spin subspace)
proved that the equilibrium spin current is the diamagnetic response to the
effective non-Abelian spin-orbit magnetic field.\cite{Tokatly08} If the spin-orbit field
is a pure gauge and, thus, can be removed by a gauge transformation, the effective $SU(2)$ magnetic field is zero, and the
equilibrium spin current vanishes.

Here we present a theory based on the gauge transformation, for spin
dynamics in a two-dimensional electron gas in the case when the spin-orbit
field can be completely removed by such a gauge transformation. We show
that the absence of the equilibrium spin current is directly related to the
giant anisotropy in the spin relaxation rate, when the relaxation does not
occur for certain spin directions.\cite{Averkiev99,Schliemann03} After gauging away
the spin-orbit coupling, the entire nonequilibrium dynamics of a {\it transformed} spin becomes almost trivial and can be described
phenomenologically exactly by only two transport coefficients which can be
determined experimentally, or calculated theretically to any desired level of accuracy. 
The first is the spin diffusion coefficient and the second is
the electron mobility required only when a constant electric field is
applied. With the inverse transformation to the initial dynamical variables, we
recover the full nontrivial dynamics of the {\it physical} spin, including the absence of the spin relaxation for
certain spin directions, that is a strong anisotropy in the spin relaxation,
stable spin configurations forming persistent spin helices, and spin precession
due to a charge current in a constant external electric field. In
addition, this approach allows making predictions for more general cases of spin-orbit
coupling, including nonuniform spin-orbit fields.

\section{Spin current and spin relaxation: the conventional approach.}

We begin with the conventional Hamiltonian of spin-orbit coupling in
two-dimensional electron gas: 
\begin{equation}
H_{\rm so}=\frac{1}{2}\sum_{j}\left( \alpha _{ja}({\bm \rho })k_{j}+k_{j}\alpha
_{ja}({\bm \rho })\right) \sigma ^{a},
\label{Hso}
\end{equation}
where $\alpha _{ja}$ is the coordinate-dependent spin-orbit coupling field, $%
k_{j}=-i\partial /\partial x_{j}$ is the momentum operator, Cartesian
subscript indices $j=x,y$ correspond to the electron coordinate ${\bm \rho }=\left(
x,y\right) ,$ and $\sigma ^{a}$ are the Pauli matrices with the upper
Cartesian indices corresponding to three directions $x,y,z$ in
the spin subspace. We use the system of units with $\hbar \equiv 1$
and sum up over repeating indices. 
Interaction $H_{\rm so}$ arises in a two-dimensional electron gas from various
sources. Two origins are considered as the most important. The first one,
arising due to the inversion asymmetry of the crystal unit cell, is
described by the Dresselhaus model. The other one is the Rashba field,\cite{Rashba84} where
the coupling originates from the macroscopic asymmetry of the structure
hosting the two-dimensional electron gas.\cite{Winkler03} Due to various physical 
origins, including material, structure, doping, and 
possible mechanical strain, numerical values of parameters $\alpha _{ja}$
vary strongly from system to system ranging from $10^{-12}$ eV$\cdot $cm for
Si- to $10^{-9}$ eV$\cdot $cm for GaAs-based structures and will not be discussed
here.

For the coordinate-independent spin-orbit field $k_{j}\alpha _{ja}({\bm \rho 
})=0$, and the Hamiltonian (\ref{Hso}) can be presented as:

\begin{equation}
H_{\rm so}=\sum_{j}\alpha _{j}\left( \mathbf{h}_{[j]}\cdot{\bm \sigma} \right) k_{j},
\label{Hso1}
\end{equation}
where $\mathbf{h}_{[j]}$ is a unit-length vector, and $\alpha _{j}$ is the
corresponding spin-orbit coupling constant for given component of momentum;
its contribution to the Hamiltonian is, therefore, proportional to the spin projection onto the $%
{\mathbf h}_{[j]}$ axis. The coupling leads to a momentum-dependent spin splitting
of the electron states. As a results, the Fermi line of the electron gas
becomes spin-dependent and two Fermi lines in the electron gas appear. The $H_{\rm so}$
Hamiltonian makes the velocity spin-dependent: 
\begin{equation}
v_{j}=\frac{k_{j}}{m}+i\left[ H_{\rm so},\rho _{j}\right] 
=\frac{k_{j}}{m}+\alpha _{j}\left( \mathbf{h}_{[j]}\cdot{\bm \sigma} \right),
\label{vj}
\end{equation}
with $m$ being the electron effective mass.

With the spin-dependent velocity in Eq.(\ref{vj}) we define the operator of the spin current
in the form:

\begin{equation}
J_{j}^{a}=\frac{1}{2}\sum_{\mathbf{k}}C_{\mathbf{k}}^{\dagger }
\left(v_{j}\tau ^{a}+\tau ^{a}v_{j}\right) C_{\mathbf{k}},
\end{equation}
where the $SU(2)$ group generators $\tau ^{a}=\sigma ^{a}/2,$ and $C_{%
\mathbf{k}}^{\dagger },C_{\mathbf{k}}$ are the corresponding spinors. 
For example, let $\left|\Phi\right>$ be the ground state wave function. The resulting expectation
value of the total spin current, summed up over all electrons in the gas is: 
\begin{equation}
\left\langle J_{j}^{a}\right\rangle
=\left\langle \Phi \right|J_{j}^{a}\left| \Phi \right\rangle.
\end{equation}
In the absence of special symmetry relations between the components of the
Hamiltonian $H_{\rm so},$ the expectation values of spin current $\left\langle
J_{j}^{a}\right\rangle $ are not zero, leading to the puzzling equilibrium
spin current without any measurable spin transport. Therefore, spin current
can be a characteristic of the equilibrium states of two-dimensional
electron gas. In the conventional calculation of $%
\left\langle J_{j}^{a}\right\rangle $ due to the spin-doubling of
the Fermi line, contributions to $\left\langle J_{j}^{a}\right\rangle $ come
from two subsystems: single- and double occupied states at a given electron
momentum. These two contributions have opposite signs and almost compensate
each other, yielding the results in the third order of the coupling constants 
$\alpha _{j}^{3}$. This third-power dependence is expected from perturbation
theory: $\left\langle J_{j}^{a}\right\rangle $ should be an odd function of
the spin-orbit coupling and vanish in the first order since in the ground
state without spin-orbit coupling the Fermi-line is not spin-split and all
states with given $\mathbf{k}$ are doubly occupied.

Another important feature of the electron gas with spin-orbit coupling is
the spin relaxation. Assume that one has initially produced a nonequilibrium
state $\Phi _{S}$ of a uniform spin density with the components: 
\begin{equation}
S^{a}=\left\langle\Phi_{S}\right|\sum_{\mathbf{k}}C_{\mathbf{k}}^{\dagger}\tau^{a}C_{\mathbf{k}}\left|\Phi_{S}\right\rangle .
\end{equation}
Then, the state $\left|\Phi_{S}\right\rangle$ will relax to the
equilibrium through all possible interactions and spin-orbit coupling. The
first stage of the process, the momentum relaxation, is fast.
If the spin-orbit coupling is weak compared to the random 
interactions causing the momentum relaxation, as it is assumed for the
rest of this paper, the following spin relaxation is relatively slow. 
As a result, at the second stage 
the spin components decrease with relaxation rates described by a
symmetric tensor $\Gamma^{ab}$: 
\begin{equation}
\frac{dS^{a}}{dt}=-\Gamma^{ab}S^{b}.
\end{equation}
The components of $\Gamma ^{ab}$ depend on the
spin-orbit coupling and all possible interactions of electrons with disorder,
phonons, and other electrons in the system.\cite{Ivchenko,Wu} If the spin-orbit coupling vanishes,
$\Gamma^{ab}=0$. 

\section{Spin-orbit coupling as a gauge field: pure gauge.}

Now we write the Hamiltonian of two-dimensional electron gas in the presence
of spin-orbit coupling as: 
\begin{equation}
\label{H}
H=\frac{1}{2m}\int dxdy\Psi ^{+}\left(i\partial_{i}+\mathcal{A}_{i}\right)^{2}\Psi + W\left(\Psi ^{+},\Psi\right)
\end{equation}
where $W\left(\Psi^{+},\Psi\right) $ contains all explicitly
spin-independent terms, including electron-electron interactions and
possibly, the effect of the external potential. The general non-Abelian
two-component potential is given by $2\times 2$ matrices: 
\begin{equation}
\mathcal{A}_{j}\equiv A_{j}^{a}\tau^{a}=
2m\alpha _{j}({\bm\rho})h_{[j]}^{a}({\bm\rho})\tau^{a}.
\label{Aj}
\end{equation}
The expression (\ref{Aj}) is valid for any arbitrary nonuniform spin-orbit coupling. 
Let us now perform at a given ${\bm\rho-}$point a local $SU(2)-$rotation \cite{Tokatly08} 
in the spin subspace by 
\begin{equation}
\label{U}
\mathbf{U}=\exp[i\theta^{a}({\bm\rho})\tau^{a}],
\end{equation}
with the transformation of the field operators: 
\begin{equation}
\label{tildePsi}
\widetilde{\Psi }^{+}\mathbf{U}^{-1}=\Psi ^{+},\qquad 
\widetilde{\Psi }=\mathbf{U}\Psi .
\end{equation}
This transformation renders the spin-independent quantities such as the
charge density and the charge current density, invariant. In contrast, the spin density
operators, 
\begin{equation}
\mathcal{S}=S^{a}\tau^{a}, 
\end{equation}
transforming as 
\begin{equation}
\mathcal{S}=\mathbf{U}\widetilde{\mathcal{S}}\mathbf{U}^{-1},
\label{transform1}
\end{equation}
exemplify covariant observable quantities. This difference between the physical
quantities which transform invariantly and covariantly under a local $SU(2)$ rotation is crucial
for the understanding of the spin dynamics. 

For the matrix 
$\mathbf{U}=\exp\left[i\theta\left(\mathbf{h}\cdot{\bm\tau}\right)\right],$ where $\mathbf{%
h}$ is a unit length vector, the  $\tau ^{b}-$matrices,  transformed according to Eq.(\ref{transform1}), 
acquire the form: 
\begin{equation}
\widetilde{\tau }^{b}= h^{b}\left(\mathbf{h}\cdot{\bm\tau}\right)+
\cos \theta [\tau ^{b}-h^{b}\left(\mathbf{h}\cdot{\bm\tau}\right)]+
\sin \theta \varepsilon^{abc}h^{a}\tau^{c},
\label{matr_transform}
\end{equation}
where $\varepsilon ^{abc}$ is the Levi-Civita tensor. This equation shows that the product 
$\mathbf{h}\cdot{\bm\tau }$ is unaffected by the 
transformation (\ref{U}). Therefore, if we present $\mathcal{S}$ as the sum of
longitudinal and transverse components 
$\mathcal{S=S}_{\parallel }\mathcal{+S}_{\perp }$ with $\mathcal{S}_{\parallel}=\mathbf{h}\left(\mathcal{S}\cdot\mathbf{h}\right) ,$ the longitudinal component (spin projection onto the $%
\mathbf{h-}$axis) remains constant, while the $\mathcal{S}_{\perp }$ does not; it rotates by the angle $\theta$ around the $\mathbf{h-}$axis).
This simple observation will be important for the further analysis in this paper.

The Hamiltonian preserves its form under a local $SU(2)$ rotation of the fermionic fields if the vector-potential is transformed as follows 
\begin{equation}
\label{tildeA}
\widetilde{\mathcal{A}}_{i}=
\mathbf{U}^{-1}\left(i\partial_{i}\mathbf{U}\right)+\mathbf{U}^{-1}\mathcal{A}_{i}\mathbf{U}.
\end{equation}
Indeed, after the transformation the Hamiltonian acquires the form:
\begin{equation}
H=\frac{1}{2m}\int dxdy\widetilde{\Psi }^{+}
\left(i\partial_{i}+\widetilde{\mathcal{A}}_{i}\right)^{2}\widetilde{\Psi }
+W\left( \widetilde{\Psi }^{+},\widetilde{\Psi}\right) ,
\end{equation}
which is identical to that of Eq.~(\ref{H}), but with $\Psi$ and $\mathcal{A}_{i}$ being replaced by the transformed quantities, $\widetilde{\Psi}$ and $\widetilde{\mathcal{A}}_{i}$, respectively.

Assume now that $\mathcal{A}_{i}$ in the original Hamiltonian (\ref{H}) corresponds to a pure gauge vector-potential, that is both $\mathcal{A}%
_{x}$ and $\mathcal{A}_{y}$ can be removed by the above transformation such
that $\widetilde{\mathcal{A}}_{x}=\widetilde{\mathcal{A}}_{y}=0$. In this
case there exists a local rotation determined by three coordinate-dependent
functions $\theta _{\mathcal{A}}^{a}(x,y)$:
\begin{equation}
\mathbf{U}_{\mathcal{A}}=\exp[i\theta _{\mathcal{A}}^{a}({\bm \rho })\tau ^{a}],
\end{equation}
such that the initial components $\mathcal{A}_{i}$ can be presented in the
form
\begin{equation}
\mathcal{A}_{i}=\mathbf{U}_{\mathcal{A}}
\left(i\partial_{i}\mathbf{U}_{\mathcal{A}}^{-1}\right) .
\end{equation}
A vector-potential of this form is gauged away by the transformation (\ref{U}) with $\mathbf{U}=\mathbf{U}_{\mathcal{A}}$: 
\begin{equation}
\widetilde{\mathcal{A}}_{i}=\mathbf{U}_{\mathcal{A}}^{-1}
\left( i\partial_{i}+\mathbf{U}_{\mathcal{A}}\left(i\partial_{i}\mathbf{U}_{\mathcal{A}}^{-1}\right)
\right) \mathbf{U}_{\mathcal{A}}=0.
\end{equation}
If the spin-orbit field can be removed by a gauge transformation,
the subsequent spin dynamics is simplified considerably and in certain regimes, 
like the drift-diffusion processes considered below, the problem becomes elementary. The inverse $SU(2)-$rotation transforms the spin
components to the actual values and we recover the full dynamics of the physical spin. 
We will follow this procedure in the present paper.

We mention a textbook example of a similar approach. 
When the motion of a relativistic electron in static perpendicular
electric field $\mathbf{E}$ and magnetic field $\mathbf{H}$ is considered,
there exists a reference frame, where, after the Lorentz transformation, the
smaller of these fields vanishes. In this frame the equations of the electron
motion are very simple, and in the case $H<E,$ they are essentially, one-dimensional.
The inverse Lorentz transformation provides the full description of the
electron motion in the presence of both fields.\cite{Jackson}

In the pure gauge field after the local $SU(2)$ transformation 
$\mathbf{U}_{\mathcal{A}}=\exp[i\theta_{\mathcal{A}}^{a}({\bm \rho })\tau ^{a}]$ the Hamiltonian takes the
form: 
\begin{equation}
H=-\frac{1}{2m}\int dxdy\widetilde{\Psi }^{+}\Delta \widetilde{\Psi }%
+W\left( \widetilde{\Psi }^{+},\widetilde{\Psi }\right),
\end{equation}
with no spin-orbit coupling present. As mentioned above, the spin dynamics
with this Hamiltonian can be formulated in general terms phenomenologically
and then by inverse transformation, returned to the form where the coupling
and full spin dynamics are restored.

Vector-potential is a pure gauge, allowing removal six terms in $\mathcal{A%
}_{x},\mathcal{A}_{y},$ with the transformation $\mathbf{U}_{\mathcal{A}}$
based on the three functions $\theta _{\mathcal{A}}^{a}({\bm \rho })$ given
certain relations between the $\mathcal{A}_{x}$ and $\mathcal{A}_{y}$
components. The corresponding conditions are naturally formulated in terms of a
non-Abelian field strength tensor $\mathcal{F}_{ij}$ : the vector potential is locally a pure gauge if the field strength is zero, 
\begin{equation}
\mathcal{F}_{ij}=\partial _{i}\mathcal{A}_{j}-\partial _{j}\mathcal{A}_{i}-
i\left[ \mathcal{A}_{i},\mathcal{A}_{j}\right] =0.
\end{equation}
\begin{figure}[tbp]
\begin{center}
\vspace*{-0.2cm} \epsfxsize=6.5cm \epsfbox{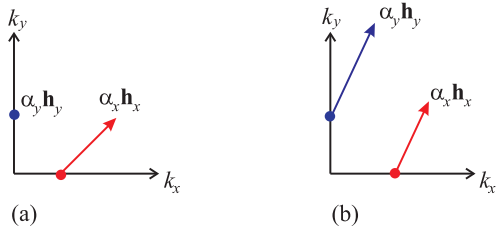} \vspace*{-0.2cm}
\end{center}
\caption{(Color online) Illustration of two cases of the pure gauge
spin-orbit field. (a) one of the coupling constants $\alpha_j$ is zero, case (i) (b)
both coupling constants are not zero, the directions of corresponding
magnetic fields coincide, case (ii). The direction of the spin-orbit field remains
constant for any electron momentum ${\bf k}=(k_x,k_y)$. }
\end{figure}
For the spatially uniform case, using Eq.(\ref{Aj}) this condition is reduced to 
$\left[ \mathcal{A}_{i},\mathcal{A}_{j}\right] =0,$ that is:

(i) either $\alpha _{i}\alpha _{j}=0,$ or

(ii) $\left[ \mathbf{h}_{[i]}\cdot{\bm \tau },\mathbf{h}_{[j]}\cdot{\bm \tau }\right]
=0$ if $\alpha _{i}\alpha _{j}\neq 0.$

The commutation relation 
\begin{equation}
\left[ \mathbf{h}_{[i]}{\bm \tau },\mathbf{h}_{[j]}{\bm \tau }\right] =
i{\bm \tau }\cdot
\left(\mathbf{h}_{[i]}\times \mathbf{h}_{[j]}\right),
\label{commutator}
\end{equation}
demonstrates that the spin projections commute only for the same axis, that is $%
\mathbf{h}_{[i]}=\pm \mathbf{h}_{[j]}$. Therefore, the solution to Eq. (\ref{commutator}) has
the form (we assume below $\mathbf{h}_{[i]}=\mathbf{h}_{[j]}$ in the case
(ii) for definiteness): 
\begin{equation}
\mathcal{A}_{j}=2m\alpha \nu _{j}\left( \mathbf{h}\cdot{\bm \tau }\right) 
\end{equation}
where $\mathbf{h}=$ $\mathbf{h}_{[i]}=\mathbf{h}_{[j]}$ if $\alpha
_{i}\alpha _{j}\neq 0$ or $\mathbf{h}=$ $\mathbf{h}_{[f]}$ for nonzero $%
\alpha _{f}$, where $f=x$ or $f=y$, as illustrated in Fig.(1). Here \textbf{%
\ }$\alpha =\left( \alpha _{x}^{2}+\alpha _{y}^{2}\right) ^{1/2},$ and ${\bm \nu 
}$ is a unit vector. The corresponding gauge transformation is: 
\begin{equation}
\mathbf{U}_{\mathcal{A}}={\exp }\left[ 2im\alpha \rho _{j}\nu _{j}\left( 
\mathbf{h}\cdot{\bm \tau }\right) \right] 
{\exp }\left[ 2im\alpha \rho _{i}\nu_{i}\left( \mathbf{h}\cdot{\bm \tau }\right) \right]
=\exp \left[ 2im\alpha\left( \mathbf{h}\cdot{\bm \tau}\right) \left({\bm \rho}\cdot{\bm\nu}\right) 
\right].
\label{Uuniform}
\end{equation}
From this condition we immediately conclude that the
projection of the total spin at the $\mathbf{h}_{[i]}=\mathbf{h}_{[j]}$ axis commutes with $%
H_{\rm so}$, and, therefore, remains constant with time for
arbitrary dynamics. Experimentally, this fact corresponds to the
vanishing relaxation for this spin direction; this conclusion crucial for
the design and application of spin-based devices. If $\alpha_{i}\alpha
_{j}=0,$ the problem immediately becomes one-dimensional, trivial from
the diamagnetic response interpretation of the equilibrium spin current, \cite{Tokatly08} 
since one-dimensional systems do not demonstrate this kind of response.
The same situation occurs in quantum wires, where the motion of electrons 
is strictly one-dimensional, no
equilibrium spin current exists, and the spin projection along the 
$\mathbf{h}_{[f]}$ axis is conserved. In the Appendix, for illustration, we perform a conventional
calculation of the equilibrium spin current in a two-dimensional electron
gas with the pure gauge spin-orbit coupling  and in a quantum wire, and demonstrate that
the spin current vanishes in both systems.

\begin{figure}[tbp]
\begin{center}
\vspace*{-0.2cm} \epsfxsize=6.5cm \epsfbox{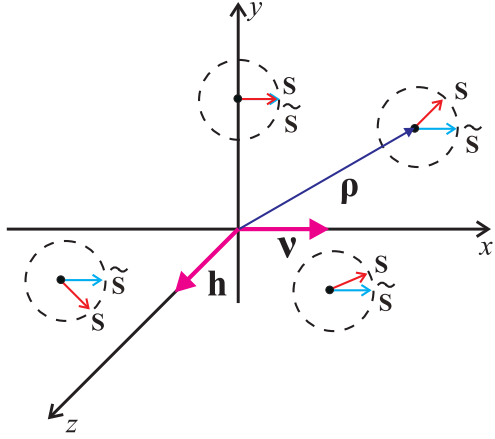} \vspace*{-0.2cm}
\end{center}
\caption{(Color online) Illustration of coordinate-dependent mutual orientation of
$\widetilde{\mathbf{S}}$ and $\mathbf{S}$ vectors corresponding to Eq.(\ref{taus:011})
for a structure grown along the [110] crystal axis. Vectors ${\bf h}$ and ${\bm \nu}$ are shown in the Figure. 
The angle between  $\widetilde{\mathbf{S}}$ and $\mathbf{S}$ is determined solely by the $x$-component
of ${\bm \rho}-$vector.}
\end{figure}

There are two widely studied
realizations of the above discussed pure gauge field. 
The $\alpha_{i}\alpha_{j}=0$ case is realized for the Dresselhaus model for the electron
gas confined in the quantum wells of GaAs grown along the $[110]$ crystal
axis. The coupling constant $\alpha$ in this system \cite{Dyakonov86} is approximately
inversely proportional to the square of the quantum well width $w$.
In this case the vector-potential and the corresponding
transformations are:
\begin{equation}
\left( \mathcal{A}_{x},\mathcal{A}_{y}\right) =
\left( 2m\alpha \tau^{z},0\right) ,\qquad 
\mathbf{U}_{\mathcal{A}}=\exp \left[2im\alpha x\tau^{z}\right],
\end{equation}
where the $z-$ axis is oriented along the growth direction and the $x$-axis
is that of the unit cell.  
Here we use transformation (\ref{Uuniform}) with $\mathbf{h=}(0,0,1),$  ${\bm \nu}=(1,0),$ and $\theta
(x,y)=2m\alpha x$ to obtain:
\begin{equation}
\widetilde{\tau }^{z}=\tau ^{z},
\qquad \widetilde{\tau }^{x}=\cos \theta\tau ^{x}+\sin \theta \tau ^{y},
\qquad \widetilde{\tau }^{y}=\cos \theta\tau ^{y}-\sin \theta \tau ^{x}.  
\label{taus:011}
\end{equation}
We illustrate the resulting relations between $\mathbf{S}$ and $\widetilde{%
\mathbf{S}}$ for this simple situation in Fig.(2): when $\widetilde{\mathbf{S%
}}$ remains constant in space, $\mathbf{S}$ turns by the angle $\theta (x,y) 
$ around the $z-$axis.

The $\alpha _{i}\alpha _{j}\neq 0$ case is realized in the compensated
Dresselhaus-Rashba model for the GaAs structure grown along the $[001]$ crystal axis.
Here
\begin{eqnarray}
\left(\mathcal{A}_{x},\mathcal{A}_{y}\right)  &=&
\left( 2m\alpha \left(\tau ^{x}-\tau ^{y}\right) ,2m\alpha \left( \tau ^{x}-\tau ^{y}\right)\right),\\
\mathbf{U}_{\mathcal{A}} &=&
\exp \left[ 2im\alpha \left( x+y\right)\left(\tau ^{x}-\tau ^{y}\right) \right] .
\label{trans011}
\end{eqnarray}
Here  we obtain with $\mathbf{h=}(1,-1,0)/\sqrt{2},$  ${\bm \nu}=(1,1)/\sqrt{2},$ 
and $\theta =2\sqrt{2}m\alpha\left( x+y\right)$: 
\begin{eqnarray}
\widetilde{\tau }^{z} &=&\cos \theta \tau ^{z}-\frac{1}{\sqrt{2}}\sin \theta
\left( \tau ^{x}+\tau ^{y}\right), \label{taus:001} \\
\widetilde{\tau }^{x} &=&\cos^2\frac{\theta}{2}\tau^{x} - \sin ^{2}\frac{\theta }{2}\tau ^{y} 
+\frac{1}{\sqrt{2}}\sin \theta \tau ^{z},
\qquad 
\widetilde{\tau }^{y}=\cos^2\frac{\theta}{2}\tau^{y} - \sin ^{2}\frac{\theta }{2}\tau^{x} 
+\frac{1}{\sqrt{2}}\sin \theta \tau ^{z}
.  \nonumber
\end{eqnarray}
Equations (\ref{taus:011}), (\ref{taus:001}) illustrate a general feature of the relations between
the original $\mathbf{S}=(S^{x},S^{y},S^{z})$ and gauge-transformed $\widetilde{\mathbf{S}}$ 
spin densities and \textit{vice versa}. For the spin-orbit field
characterized by the direction $\mathbf{h,}$ for a uniform
coordinate-independent $\mathbf{S,}$ the components $\mathbf{S}_{\parallel }$
and $\widetilde{\mathbf{S}}_{\parallel }$ coincide. The $\widetilde{\mathbf{S}}_{\perp }$-component 
forms a periodic structure on the spatial scale of the
order of $L_{\rm so}=1/m\alpha$, or $\hbar ^{2}/m\alpha $ when the units are
restored.  The mean value 
$\left\langle \widetilde{\mathbf{S}}_{\perp}(x,y)\right\rangle =0$ for the infinitely
large systems considered here, where the boundary conditions do not change the spin
dynamics. The meaning of the length $L_{\rm so}$ can be understood as
follows. Hamiltonian (\ref{Hso1}) shows that the spin-orbit
coupling $H_{\rm so}$ causes for an electron moving with the velocity ${\bf v}$,  
spin precession around $\mathbf{h}$  with the rate of the order of $\alpha mv.$ The
corresponding precession angle is of the order of $\alpha mL,$ where $L=vt$ is the electron
displacement. Thefore, $L_{\rm so}$ can be viewed as the travel distance at which
the electron spin can undergo a full rotation. Another circumstance is,
however, more important: the spin rotation angle depends only on the electron
displacement and not on the details of its motion between initial and final
points, leading to the appearance of stable spin structures, discussed below. Here
a numerical value of typical $L_{\rm so}$ can be of interest. For GaAs with $%
m=0.067m_{0},$ where $m_{0}$ is the free electron mass, and $\alpha $ of the
order of $10^{-7}$ meV$\cdot$cm, $L_{\rm so}$ is of the order of several microns.

In both these systems, the observed spin relaxation rate is strongly
anisotropic with one spin component having lifetime orders of magnitude
longer than the others. The weak relaxation rate for these components is
determined by the mechanisms different from the homogeneous spin-orbit
coupling, most probably, related to the disorder in the spin-orbit
coupling. \cite{Muller08,Glazov05,Dugaev09}

\section{Spin dynamics: diffusion, precessional behavior, and drift
contributions.}

\begin{figure}[tbp]
\begin{center}
\vspace*{-0.2cm} \epsfxsize=6.5cm \epsfbox{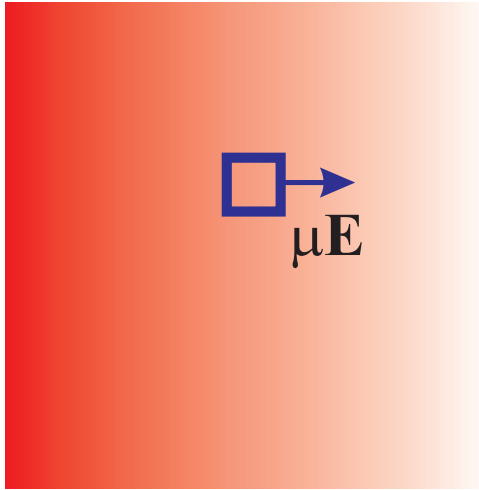} \vspace*{-0.2cm}
\end{center}
\caption{(Color online) Nonuniform spin density evolving by
diffusion and drift dynamics. Small square with an arrow illustrates the
effect of the external electric field on the nonuniform spin dynamics.}
\end{figure}

After the gauge transformation, the spin-orbit interactions is switched off.
Therefore, on a time scale much longer than the momentum relaxation
time, the spin dynamics becomes combination of pure spin diffusion and spin
drift: 
\begin{equation}
\label{tilde-diff}
\partial _{t}\widetilde{\mathcal{S}}=
D\Delta \widetilde{\mathcal{S}}+
\mu E_{j}\partial_{j}\widetilde{\mathcal{S}},
\end{equation}
where $D$ is the spin-diffusion coefficient, $\mu $ is the electron
mobility, and $\mathbf{E}$ is the two-dimensional
applied electric field \cite{Amico01} as illustrated in Fig.(3). 
In Eq.(\ref{tilde-diff}) we have taken into account that the uniform velocity of electrons 
is $-\mu\mathbf{E}$. These two parameters fully
describe the drift-diffusive spin dynamics in the absence of spin-orbit
coupling. Macroscopic motion of electrons (electric currents) can drag nonuniform spin
density between different parts of the electron gas. This effect leads to
the $\mu E_{j}\partial_{j}\widetilde{\mathcal{S}}$ term in $\partial
_{t}\widetilde{S}$. The initial spin density 
eventually vanishes due to diffusion, however, the total integrated spin polarization will remain constant.
The diffusive evolution of the transformed spin density $D\Delta 
\widetilde{\mathcal{S}}$ occurs if the electron free path of the order of
$\ell =v\tau_{p}$  is much less than
the spatial scale of the inhomogeneity: $\ell\ll L_{\rm so}.$ This condition can
be formulated as $\Omega_{\rm so}\tau_{p}\ll 1,$ meaning that the spin-orbit
coupling is relatively weak. The spatial inhomogeneity of the order of 
$L_{\rm so}$ and $D$ of the order of $v^{2}\tau _{p}$ set the 
time scale of the diffusion  smearing of the $\widetilde{\mathcal{S}}$ as 
$\widetilde{t}_{D}\sim L_{\rm so}^{2}/D$ on the order of $\Omega _{\rm so}^{-2}\tau_{p}^{-1},$
and, therefore, the same spin relaxation time for real spin $\mathcal{S}.$

The evolution of the physical measurable spin density: 
\begin{equation}
\label{inverse-transf}
\mathcal{S}=\mathbf{U}_{\mathcal{A}}\widetilde{\mathcal{S}}\mathbf{U}_{\mathcal{A}}^{-1}, 
\end{equation}
is due to the diffusion and spin precession since the transition of electron
from point ${\bm \rho}_{1}$ to point ${\bm \rho}_{2}$ is accompanied by the rotation of
its spin, dependent only on the displacement ${\bm \rho}_{2}-{\bm \rho}_{1}$. 
Irregular motion in the diffusion process is seen in the spin
relaxation, and regular drift causes spin precession, with these two
processes being mutually related.

Motion of $\mathcal{S}$ is described, therefore, by the following equations for the time
evolutions of the spin density, which are obtained by applying the inverse transformation (\ref{inverse-transf}) to the drift-diffusion equation (\ref{tilde-diff}), 
\begin{equation}
\partial _{t}\mathcal{S}=
D\mathbf{U}_{\mathcal{A}}\left[ \Delta \left( 
\mathbf{U}_{\mathcal{A}}^{-1}\mathcal{S}\mathbf{U}_{\mathcal{A}}\right)
\right]\mathbf{U}_{\mathcal{A}}^{-1}
+\mu E_{j}\mathbf{U}_{\mathcal{A}}
\left[ \partial _{j}\left( \mathbf{U}_{\mathcal{A}}^{-1}\mathcal{S}
\mathbf{U}_{\mathcal{A}}\right)\right]\mathbf{U}_{\mathcal{A}}^{-1}.
\end{equation}
The resulting most general equation of motion valid for any pure gauge
spin-orbit field takes the form
\begin{eqnarray}
&&\partial _{t}\mathcal{S}-D\Delta \mathcal{S}
-\mu E_{j}\partial _{j}\mathcal{S}  \nonumber \\
&=&D\left\{ 2\left[\mathbf{U}_{\mathcal{A}}\Bbb{\nabla }\mathbf{U}_{%
\mathcal{A}}^{-1},\Bbb{\nabla }\mathcal{S}\right] -
2\left( \mathbf{U}_{\mathcal{A}}\Bbb{\nabla }\mathbf{U}_{\mathcal{A}}^{-1}\right) \mathcal{S}%
\left( \mathbf{U}_{\mathcal{A}}\Bbb{\nabla }\mathbf{U}_{\mathcal{A}}^{-1}\right) 
+\left( \mathbf{U}_{\mathcal{A}}\Delta \mathbf{U}_{\mathcal{A}}^{-1}\right) \mathcal{S}
+\mathcal{S}\left( \Delta \mathbf{U}_{\mathcal{A}}\right)\mathbf{U}_{\mathcal{A}}^{-1}\right\}   \nonumber \\
&&+\mu E_{j}\left[ \left( \mathbf{U}_{\mathcal{A}}\partial_{j}\mathbf{U}_{\mathcal{A}}^{-1}\right) ,\mathcal{S}\right] .
\end{eqnarray}
The total expression for local evolution of spin density components can be
obtained from this equation by multiplying both sides by $\tau ^{a}$ and
taking the trace using the identity: $\mathrm{tr}\left( \tau
^{a}\tau ^{b}\right) =\delta ^{ab}/2.$ The result can be presented as the sum:

\begin{equation}
\partial _{t}S^{a}=D\Delta S^{a}+
\mu E_{j}\partial_{j}S^{a}+B_{j}^{ab}\partial _{j}S^{b}-H^{ab}S^{b}-\Gamma ^{ab}S^{b}.
\end{equation}
The general expressions for non-diagonal tensors of kinetic coefficients entering this equation are: 
\begin{eqnarray}
B_{j}^{ab} &=&-B_{j}^{ba}=4D\mathrm{tr}
\left\{
\tau^{a}\left[\mathbf{U}_{\mathcal{A}}\partial _{j}\mathbf{U}_{\mathcal{A}}^{-1},\tau ^{b}\right]
\right\},  \\
H^{ab} &=&-H^{ba}=
2\mu E_{j}\mathrm{tr}
\left\{ 
\tau ^{a}\left[ \mathbf{U}_{\mathcal{A}}\partial _{j}\mathbf{U}_{\mathcal{A}}^{-1},\tau ^{b}\right]
\right\},  \\
\Gamma ^{ab} &=&4D\left( \mathrm{tr}\left\{ \tau ^{a}\left( \mathbf{U}_{%
\mathcal{A}}\partial _{j}\mathbf{U}_{\mathcal{A}}^{-1}\right) \tau
^{b}\left( \mathbf{U}_{\mathcal{A}}\partial _{j}
\mathbf{U}_{\mathcal{A}}^{-1}\right) \right\} -
\frac{1}{2}\mathrm{tr}\left\{\tau^{a}\left( 
\mathbf{U}_{\mathcal{A}}\Delta \mathbf{U}_{\mathcal{A}}^{-1}\right) \tau
^{b}+\tau ^{a}\left( \Delta \mathbf{U}_{\mathcal{A}}
\right)
\mathbf{U}_{\mathcal{A}}^{-1}\tau ^{b}\right\} \right)   \nonumber \\
&&
\end{eqnarray}
Now we can study the physical meaning of the obtained non-diagonal tensors and
simplify the expressions for the time derivatives for uniform
spin-orbit coupling with: 
\begin{equation}
\mathbf{U}_{\mathcal{A}}=\exp 
\left[ 2im\alpha \left( \mathbf{h}\cdot{\bm \tau }\right) ({\bm \nu}\cdot{\bm \rho}) \right] ,
\qquad \mathbf{U}_{\mathcal{A}}^{-1}=\mathbf{U}_{\mathcal{A}}^{+}
=\exp \left[ -2im\alpha \left( \mathbf{h}{\bm \tau }\right)
({\bm \nu}\cdot{\bm \rho}) \right],
\label{UA}
\end{equation}
With the given form in Eq.(\ref{UA}) of $\mathbf{U}_{\mathcal{A}}$ and $\mathbf{U}_{\mathcal{A%
}}^{-1}$ we obtain: 
\begin{equation}
\mathbf{U}_{\mathcal{A}}\Bbb{\nabla }\mathbf{U}_{\mathcal{A}}^{-1}=
-2im\alpha \left( \mathbf{h}\cdot{\bm \tau }\right) {\bm \nu ,}
\qquad
\Delta \mathbf{U}_{\mathcal{A}}=-4m^{2}\alpha ^{2}\mathbf{U_{\mathcal{A}}},
\qquad 
\Delta \mathbf{U}^{-1}=-4m^{2}\alpha ^{2}\mathbf{U}_{\mathcal{A}}^{-1}.
\label{UAdiff}
\end{equation}
With formulas (\ref{UA}),(\ref{UAdiff}) we obtain for the diffusion-related coefficients: 
\begin{eqnarray}
B_{j}^{ab} &=&-2m\alpha _{j}D\varepsilon ^{abc}h^{c}, \\
\Gamma ^{ab} &=&4m^{2}\alpha ^{2}D\left( \delta ^{ab}-h^{a}h^{b}\right) .
\end{eqnarray}
The corresponding drift-dependent contribution:
\begin{equation}
H^{ab}=-m\alpha \mu \left( {\bm \nu }\cdot \mathbf{E}\right) \varepsilon
^{abc}h^{c},
\end{equation}
describes the spin precession.

The answer for the spin density  $\mathbf{S}$ with components $\left(
S^{x},S^{y},S^{z}\right)$ has the form of three terms of different order
in $\alpha $: 
\begin{equation}
\label{diff}
\partial _{t}\mathbf{S}=\left. \partial _{t}\mathbf{S}\right| _{0}+
\left.\partial _{t}\mathbf{S}\right| _{1}+
\left.\partial_{t}\mathbf{S}\right|_{2}.
\end{equation}
These terms have different meaning and can be expressed as: 
\begin{eqnarray}
\left. \partial _{t}\mathbf{S}\right| _{0} &=&
D\Delta \mathbf{S}+\mu E_{j}\partial _{j}\mathbf{S},  \label{dsdt1} \\
\left. \partial _{t}\mathbf{S}\right| _{1} &=&
4mD\alpha \left( {\bm\nu }\cdot \nabla \right) \left( \mathbf{h}\times \mathbf{S}\right) 
+2m\alpha \mu \left( {\bm \nu }\cdot \mathbf{E}\right) \left( \mathbf{h}\times \mathbf{S}\right),   \label{dsdt2} \\
\left. \partial _{t}\mathbf{S}\right| _{2} &=&
-4m^{2}\alpha ^{2}D\left(\mathbf{S}-\mathbf{h}(\mathbf{h}\cdot\mathbf{S})\right).  \label{dsdt3}
\end{eqnarray}
The $\left. \partial _{t}\mathbf{S}\right| _{0}$ term describes the standard drift-diffusion spin
dynamics for zero spin-orbit coupling.

The $\left. \partial _{t}\mathbf{S}\right| _{1}$ term corresponds to the spin
precession due to the spin-orbit coupling. The mobility-determined
contribution in $\left. \partial _{t}\mathbf{S}\right| _{1}$ is the
precession in the macroscopic spin-orbit field arising due to the uniform
velocity of electrons. When the electric current is induced, the momentum
distribution function is shifted such that the momentum has a nonzero value. As a
result, the Hamiltonian $H_{\rm so}$ forms a macroscopic spin-orbit Zeeman field \cite{Wilamowski08} and,
as a result, a regular spin precession $\partial_{t}\mathbf{S=}2m\alpha \mu
\left( {\bm \nu }\cdot \mathbf{E}\right) \left( \mathbf{h}\times \mathbf{S}%
\right).$  If $\left( {\bm \nu }\cdot \mathbf{E}\right)=0$, contributions
of the momentum changes along the $x$ and $y$-axes in the macroscopic spin-orbit ``magnetic''
field compensate each other, and no regular precession occurs. 
Thus, Eq.(\ref{dsdt2}) reproduces the diffusive and
non-diffusive spin precession.

The $\left.\partial_{t}\mathbf{S}\right|_{2}$ term is the Dyakonov-Perel'
mechanism of spin relaxation,\cite{Dyakonov73} which can be seen from the
fact that $D$ is determined by $\left\langle v^{2}\right\rangle \tau _{p}$,
where $v$ is the electron velocity (see, also in Ref. \cite{Raimondi}). 
Taking into account that electron momentum is $mv,$ one can see that $%
\left. \partial _{t}\mathbf{S}\right| _{2}$ corresponds to the
Dyakonov-Perel relaxation with the relaxation rate on the order of $\alpha
^{2}k^{2}\tau_p$. The obtained relation
between the spin relaxation rate and diffusion coefficient is universal. For
two different systems with the same sample-dependent $m\alpha $ parameter,
the ratio of $\Gamma ^{ab}/D$ remains constant. Since the parameters $\Gamma
^{ab}$ and $D$ can be measured independently, this universality can be
verified experimentally. For example, in the measurements performed at the
same sample at different temperatures, the ratio $\Gamma ^{ab}/D$ is
expected to remain constant for degenerated and non-degenerated electron gas.

Eqs.(\ref{dsdt2}) and (\ref{dsdt3}) show that $\mathbf{S}_{\parallel}=\mathbf{h}\cdot\mathbf{S}$ 
does not change with time,
as expected, and the entire dynamics is solely due to the $\mathbf{S}%
_{\perp }-$component. As a simple illustration we consider the evolution of an 
initially homogeneous spin density. By solving equations (\ref{diff})-(\ref{dsdt3}) 
with the initial condition ${\bf S}(\bm\rho,t=0)={\bf S}_0$ we find the spin dynamics
\begin{equation}
 \label{decay}
{\bf S}(t) = {\bf h}({\bf h}\cdot{\bf S}_0) + \left\{\cos(\Omega_Et)[{\bf S}_0-{\bf h}({\bf h}\cdot{\bf S}_0)] +
\sin(\Omega_Et) ({\bf h}\times{\bf S}_0) \right\}e^{-\Gamma t},
\end{equation}
where $\Omega_E = 2\alpha m\mu({\bm\nu}\cdot{\bf E})$ is the precession frequency in a drift-induced spin-orbit Zeeman field, and $\Gamma=4\alpha^2m^2D$ is the diffusion related relaxation rate. From Eq.~(\ref{decay}) we see that the spin precesses with the frequency $\Omega_E$ about the ${\bf h}$-axis and its transverse component decays at the rate $\Gamma$ in such a way that the projection of the spin at  ${\bf h}$ remains stationary. By comparing the characteristic time scales of the drift-induced precession and the diffusion-induced relaxation, we can estimate the external electric field at which the role of 
the drift-dependent terms becomes important; in particular, the precession becomes visible at the scale of the relaxation time. From the condition of still visible precession $\Omega_E\sim\Gamma$ we 
find the corresponding electric field $E\sim \alpha mD/\mu$. In this field, the
precession rate $\Omega_E$ is of the order of $\Omega_{\rm so}^{2}\tau_p$, making the contributions of regular
and random motion in the precession angle of the same order. 
If the spin diffusion is dominated by the impurity scattering, then $D$ and $\mu$ are proportional to
the momentum relaxation time $\tau_{p}$, and this electric field is disorder-independent. However, it
can change with the temperature since in the non-degenerated gas 
$D$ approaches the electron diffusion coefficient \cite{Amico01} and, therefore, by 
the Einstein relation $D=\mu T$. 

Another interesting effect of a spin-orbit coupling, which follows straightforwardly  from our formulation -- the existence of stable spatially inhomogeneous spin configuration. It is easy to verify that a general stationary ($\partial _{t}\mathbf{S}=\mathbf{0}$) solution to the equations (\ref{diff})-(\ref{dsdt3}) is of the form
\begin{equation}
\label{helix}
{\bf S}(\bm\rho) = {\bf h}({\bf h}\cdot{\bf S}_0) + \cos (2m\alpha(\bm{\nu}\cdot{\bm\rho}))[{\bf S}_0-{\bf h}({\bf h}\cdot{\bf S}_0)]-
\sin(2m\alpha(\bm{\nu}\cdot{\bm\rho}))({\bf h}\times{\bf S}_0),
\end{equation}
where ${\bf S}_0$ is an arbitrary constant vector. This spatially inhomogeneous stationary 
solution to the drift-diffusion equation arises due to the symmetry of the system. 
As it was demonstrated for the particular case of the model with balanced Rashba 
and Dresselhaus couplings, the symmetry can protect electron spins from 
relaxation \cite{Schliemann03} and allows for the persistent spin helix structures 
\cite{Bernevig06,Koralek09} of the form of Eq.(\ref{helix}). 
The fact that the shape of this configuration does not depend on
the diffusion coefficient $D$ shows that the persistent spin structure is insensitive
to the spin-independent disorder, in agreement with Ref.\cite{Bernevig06} 
The analysis of the spin helix stability in the presence of disordered 
spin-orbit coupling can be found in Ref.\cite{Liu06} 
It is interesting to note that the helix structure is also insensitive to the presence of the electric field and, therefore, to the mobility and presence of a transport charge current, at least in the linear Ohm's law regime. 
This seemingly counterintuitive result follows from the fact that the drift of the helix governed by the second term in (\ref{dsdt1}) is exactly compensated by the spin precession in the current-induced effective Zeeman field, the second term in (\ref{dsdt2}). A similar cancellation occurs in the diffusion channel. A diffusive spreadout of the helix, the first term in (\ref{dsdt1}), and the relaxation of the transverse component of the spin, (\ref{dsdt3}), are balanced by the ``gradient-precession`` contribution, the first term in (\ref{dsdt2}). The persistent 
spin helix configuration (\ref{helix}) has an extremely simple interpretation in terms of the transformed spin density $\widetilde{\bf S}$. The general stationary solution of the standard drift-diffusion equation (\ref{tilde-diff}) is simply a constant $\widetilde{\bf S}=\widetilde{\bf S}_0$. The relation between the physical and transformed spin density
components yield the conservation $\widetilde{\bf S}_0\cdot\mathbf{h}={\bf S}\cdot\mathbf{h}$. The
perpendicular $\widetilde{\bf S}_{0,\perp}$ is transformed according to Eq.~(\ref{inverse-transf}) 
as $\mathcal{S}=\mathbf{U}_{\mathcal{A}}\widetilde{\mathcal{S}}\mathbf{U}_{\mathcal{A}}^{-1}$ 
with $\mathbf{U}_{\mathcal{A}}=\exp\left[2im\alpha\left(\mathbf{h}\cdot{\bm \tau}\right)\left({\bm \rho}\cdot{\bm\nu}\right)\right]$
from Eq.~(\ref{Uuniform}) according to 
Eq.~(\ref{matr_transform}). The sum of the transformed terms is precisely the 
persistent spin helix of Eq.~(\ref{helix}). 

It is also instructive to look at the precession and relaxation of a spatially homogeneous spin ${\bf S}(t)$, Eq.~(\ref{decay}), from the point of view of the dynamics of the transformed spin density $\widetilde{\bf S}$. The initial condition 
${\bf S}(\bm\rho,t=0)={\bf S}_0$ for the physical spin is mapped to the initial configuration for $\widetilde{\bf S}$ in a form of a spin helix that is similar to Eq.~(\ref{helix}). The subsequent evolution of $\widetilde{\bf S}$ is governed by the standard drift-diffusion equation (\ref{tilde-diff}). Therefore the dynamical behavior is obvious -the initial helix for the transformed spin moves with the drift velocity $v_{\rm drift}=m\alpha\mu(\bm{\nu}\cdot{\bf E})$, and washes out
diffusively. When transformed to the physical spin, the drift of the helix translates to the precession, while its diffusive decay is mapped to the relaxation of the physical spin. This interpretation clearly explains why the relaxation time of the transverse components of the spin is universally determined by the diffusion coefficient. The relaxation of the physical spin components is gauge-equivalent to a purely diffusive process of washing out the initial helix configuration of the transformed spin. 

\section{Conclusions.}

We developed a gauge theory of macroscopic spin dynamics in a
two-dimensional electron gas when the spin-orbit coupling can be described
as a pure gauge, and, therefore, removed by a local $SU(2)$ rotation in the
spin subspace. We have shown that for a pure spin gauge, equilibrium spin
current vanishes and a selected axis of conserved spin projection appears
simultaneously, demonstrating gauge-related symmetry relation of these
effects. After removing the spin-orbit coupling, we considered macroscopic
phenomenological equations of spin dynamics, including spin diffusion and
spin drift in an external electric field. By the inverse $SU(2)$ rotation we
obtained the full system of partial differential equations for the time- and
spatial measured spin density evolution. This system reproduces the physics
of spin precession, stable spin configurations such as the persistent spin helix, 
and the resulting strongly anisotropic 
spin relaxation. Since we described the system without spin-orbit coupling
phenomenologically, our approach is valid at any temperature and electron 
concentration. It predicts that
the ratio of the spin relaxation rate to the spin diffusion coefficient
remains temperature- and electron concentration-independent if the coupling constants
do not depend on these two system parameters.

We presented explicit equations for the spatially uniform spin-orbit
coupling and their solutions describing  stable nonuniform structures, 
the precession and the relaxation of uniform spin polarization.  
These equations can be explicitly generalized for nonuniform
two-dimensional electron gas in macroscopic systems. We mention two of them. 
The first one  is the GaAs quantum well grown along the $\left[110\right]$  
direction with a modulated width $w(x)$, where the spin-orbit $\alpha(x)$ 
originated from the Dresselhaus coupling, varies as $1/w^{2}(x)$. 
The corresponding spin-orbit field $\mathcal{A}_{x}=2m\alpha(x)\tau^{z}$,
$\mathcal{A}_{y}=0$, with $\partial\mathcal{A}_{x}/\partial y=0$ 
remains a pure gauge. The other example is the balanced Rashba-Dresselhaus model
with the coordinate-dependent Rashba and Dresselhaus parameters 
remaining exactly equal or exactly opposite everywhere. As in the $\left[110\right]$ structure,
variation in the Dresselhaus term 
is due to the controlled variation in the structure width, while the control
of the Rashba coupling is achieved by a coordinate-dependent bias
across the well. A different kind of inhomogeneity occurs in mesoscopic systems where the 
effect of the boundary conditions for the coupled spin-charge dynamics  becomes 
important.\cite{Tserkovnyak07,Bleibaum06,Duckheim09} Generalization
of the gauge theory approach for the dynamics at the sample boundaries  
can be an interesting extension of our analysis for the infinite systems. 
Spin dynamics in these systems is of interest for the fundamental 
understanding of spin transport and for applications in spintronics devices.

\section{Acknowledgment.}

IVT acknowledges funding by the Spanish MEC (FIS2007-65702-C02-01), "Grupos
Consolidados UPV/EHU del Gobierno Vasco" (IT-319-07), and the European Community
through e-I3 ETSF project (Grant Agreement: 211956).
EYS is grateful to the  University of Basque Country UPV/EHU for support by the
Grant GIU07/40.

\section{Appendix.}

Here we show by a conventional calculation  of the spin current that
it vanishes at the equilibrium in the considered above pure gauge spin-orbit
coupling in a two-dimensional electron gas and, similarly, in one-dimensional
quantum wires. We begin with the pure-gauge two-dimensional Hamiltonian, 

\begin{equation}
H=\frac{k_{x}^{2}}{2m}+\alpha _{x}\left( h^{a}\sigma ^{a}\right) k_{x}+
\frac{k_{y}^{2}}{2m}+\alpha _{y}\left( h^{a}\sigma ^{a}\right) k_{y},
\label{H2}
\end{equation}
where $\mathbf{h}$ is unit length vector and $\alpha _{x},\alpha _{y}$ are
the corresponding spin-orbit coupling constants. The spectrum of electrons 
described by Eq.(\ref{H2}) is the sum of $k_x$ and $k_y$-dependent terms
for the two spin branches ''+`` and ''-``:
\begin{equation}
\varepsilon_{\pm }\left( k_{x},k_{y}\right) =
\frac{k_{x}^{2}+k_{y}^{2}}{2m}\pm \left( \alpha _{x}k_{x}+\alpha _{y}k_{y}\right).
\label{S2}
\end{equation}
Eq. (\ref{H2}) demonstrates that the system with pure gauge spin-orbit coupling 
remains in a certain sense, one dimensional: 
spin-orbit coupling does not couple different components of momentum in the
spectrum. For illustration we consider only the $x-$component of momentum
and velocity: 
\begin{equation}
\frac{\partial \varepsilon _{\pm }\left(\mathbf{k}\right) }{\partial k_{x}}%
=\frac{k_{x}}{m}\,\pm \alpha _{x},\qquad v_{x}=
i\left[ H,x\right] =\frac{k_{x}}{m}+\alpha _{x}\left( h^{a}\sigma ^{a}\right) ,
\end{equation}
yielding the spin current component: 
\begin{equation}
J_{x}^{b}=\frac{1}{4}
\sum_{\mathbf{k}}C_{\mathbf{k}}^{\dagger }\left\{v_{x},\sigma ^{b}\right\} C_{\mathbf{k}}.
\end{equation}
Taking into account that
$\left\{\sigma ^{a},\sigma ^{b}\right\}=2\delta^{ab}$, we obtain the anticommutator: 
\begin{equation}
\frac{1}{2}\left\{ v_{x},\sigma ^{b}\right\} =
\frac{1}{2}\left\{\frac{k_{x}}{m}+\alpha _{x}\left( h^{a}\sigma ^{a}\right),\sigma ^{b}\right\} 
=\frac{k_{x}}{m}\sigma ^{b}+\alpha _{x}h^{b}.
\end{equation}
The total spin current is the sum of contributions of two subsystems $%
\left\langle J_{x}^{b}\right\rangle =\left\langle J_{x}^{b}\right\rangle
_{+}+\left\langle J_{x}^{b}\right\rangle _{-}=2\left\langle
J_{x}^{b}\right\rangle _{+}.$ Taking into account that for given branch $%
\left\langle \sigma ^{b}\right\rangle _{\pm }=\pm h^{b}/h,$ the $%
\left\langle J_{x}^{b}\right\rangle _{+}$ spin current component becomes:
\begin{equation}
\left\langle J_{x}^{b}\right\rangle _{+}=\frac{1}{2}h^{b}\int dk_{y}\int 
\frac{\partial \varepsilon \left(\mathbf{k}\right) }{\partial k_{x}}dk_{x},
\end{equation}
where the integration is perfromed over the area in momentum space occupied
by electrons from the branch. The value of this integral is zero since this
area is restricted by the line of the constant Fermi energy $E_{F}.$ 

For one-dimensional case the situation is the same. We take the
Hamiltonian:
\begin{equation}
H=\frac{k^{2}}{2m}+\alpha \left( h^{a}\sigma ^{a}\right) k.
\end{equation}
The eigenstates of this Hamiltonian form two branches: 
\begin{equation}
\varepsilon_{\pm }=\frac{k^{2}}{2m}\pm \alpha k,
\end{equation}
corresponding to two parabolas with the minima at $-k_{0}$ and $k_{0}=\alpha
m,$ respectively, as shown in Fig.(4). 
\begin{figure}[tbp]
\begin{center}
\vspace*{-0.2cm} \epsfxsize=6.5cm \epsfbox{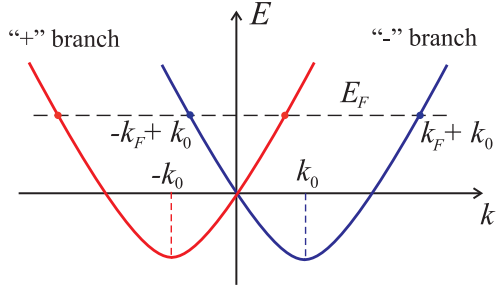} \vspace*{-0.2cm}
\end{center}
\caption{(Color online) Scheme of the spin-orbit split states in a quantum
wire. $E_F$ is the Fermi energy.}
\end{figure}
To calculate the spin current directly, we perform integration over momenta
and summation over spin branches. The ground state expectation value is: 
\begin{equation}
\left\langle J^{b}\right\rangle =
\frac{1}{2}\left[
\int_{-k_{0}-k_{F}}^{-k_{0}+k_{F}}\left( \frac{k}{m}\left\langle \sigma^{b}\right\rangle _{+}+h^{b}\right)dk+
\int_{k_{0}-k_{F}}^{k_{0}+k_{F}}\left( \frac{k}{m}\left\langle \sigma^{b}\right\rangle _{-}+h^{b}\right) dk
\right] ,
\end{equation}
where $k_{F}$ is the Fermi momentum determined by the total concentration of
electrons $n$ as $k_{F}=\pi n/2$. We obtain 
\begin{eqnarray}
\left\langle J^{b}\right\rangle  &=&2k_{F}h^{b}+\frac{1}{2m}
\left[
\int_{-k_{0}-k_{F}}^{k_{0}-k_{F}}k\frac{h^{b}}{h}dk-%
\int_{-k_{0}+k_{F}}^{k_{0}+k_{F}}k\frac{h^{b}}{h}dk
\right]   \nonumber \\
&=&2k_{F}h^{b}-\frac{2}{m}\frac{h^{b}}{h}k_{F}k_{0}.
\end{eqnarray}
The minimum position $k_{0}=\alpha m,$ yields $\left\langle
J^{b}\right\rangle =0,$ as expected.

The absence of the spin current in a wire can be related to the vanishing
persistent Aharonov-Bohm spin current \cite{Citro07,Sun07} in a ring with
spin-orbit coupling as the ring radius goes to infinity. Indeed, if a ring
is pierced by a small-radius solenoid with a finite magnetic field flux
confined inside it, the field at the ring is a pure gauge, and an
Aharonov-Bohm current in the ring arises. The spin-orbit coupling in a ring
can be understood in terms of a spin-dependent gauge, and a corresponding
spin current can be induced. This spin current vanishes in the 
$R\rightarrow\infty $ limit. In general, however, these two objects have a different topology: a ring
can demonstrate a diamagnetic response, while a wire cannot.

\end{document}